# Motion Compensated Self Supervised Deep Learning for Highly Accelerated 3D Ultrashort Echo Time Pulmonary MRI

Zachary Miller[1], Kevin M. Johnson[2,3]

1. Department of Biomedical Engineering, University of Wisconsin, Madison, Wisconsin, USA

2. Department of Medical Physics, University of Wisconsin School of Medicine and Public Health, Madison, Wisconsin, USA

3. Department of Radiology, University of Wisconsin School of Medicine and Public Health, Madison, Wisconsin, USA

Correspondence:
Kevin M. Johnson,
Department of Medical Physics, University of Wisconsin,
1133 Wisconsin Institutes for Medical Research (WIMR),
1111 Highland Ave,
Madison, WI 53705 U.S.A.
Email: kmjohnson3@wisc.edu

## Abstract:

**Purpose:** To investigate motion compensated, self-supervised, model based deep learning (MBDL) as a method to reconstruct free breathing, 3D Pulmonary ultrashort echo time (UTE) acquisitions.

**Theory and Methods**: A self-supervised eXtra Dimension MBDL architecture (XD-MBDL) was developed that combined respiratory states to reconstruct a single high-quality 3D image. Non-rigid, GPU based motion fields were incorporated into this architecture by estimating motion fields from a low resolution motion resolved (XD-GRASP) iterative reconstruction. Motion Compensated XD-MBDL was evaluated on lung UTE datasets with and without contrast and was compared to constrained reconstructions and variants of self-supervised MBDL that do not consider respiratory motion.

**Results:** Images reconstructed using XD-MBDL demonstrate improved image quality as measured by apparent SNR, CNR and visual assessment relative to self-supervised MBDL approaches that do not account for dynamic respiratory states, XD-GRASP and a recently proposed motion compensated iterative reconstruction strategy (iMoCo). Additionally, XD-MBDL reduced reconstruction time relative to both XD-GRASP and iMoCo.

**Conclusion:** A method was developed to allow self-supervised MBDL to combine multiple respiratory states to reconstruct a single image. This method was combined with GPU-based image registration to further improve reconstruction quality. This approach showed promising results reconstructing a user-selected respiratory phase from free breathing 3D pulmonary UTE acquisitions.

# 1. Introduction

3D non-Cartesian trajectories are increasingly used for free breathing MRI as they are inherently motion robust, allow for imaging of multiple respiratory states, and permit retrospective respiratory motion compensation. Such acquisitions have the potential to be applied to a range of clinical applications including ultras short echo time (UTE) lung[1–3], 4D-Flow[4,5], and Dynamic Contrast Enhanced (DCE) imaging[6]. These scans are particularly powerful when imaging patients who have difficulty breath-holding (e.g. pediatric imaging[7–9]), and when high spatial resolution imaging is needed[1]. While 3D non-Cartesian acquisitions offer intrinsic motion robustness, they have benefited substantially from advanced image reconstruction techniques such as temporal compressed sensing (e.g. XD-GRASP[10]). Such constrained reconstruction methods are computationally demanding and often require heuristic parameter tuning making these acquisitions difficult to integrate into clinical practice. Further, they often fail to provide sufficient image quality, particularly when reconstructing highly accelerated frames close to end-inspiration.

Model based deep learning (MBDL) methods applied to these acquisitions have the potential to simultaneously reduce reconstruction time, improve image quality relative to compressed sensing[11,12], and remove the need for heuristic selection of regularization transforms and parameters. These methods are similar to classical iterative approaches that alternate between data consistency steps that enforce the physical model of acquisition and regularizer steps that constrain solutions to have certain preferred properties. MBDL typically unrolls an alternating optimization for a fixed number of iterations and replaces fixed regularizers with convolutional neural networks (CNN) that learn directly from training data to remove under-sampling and other artifacts. While the application of MBDL to high spatial resolution, 3D non-Cartesian imaging has been challenged by GPU memory constraints, recent progress has been made using gradient checkpointing with patch-based methods[13] and distributed training[14].

MBDL architectures are typically trained using supervised learning methods[11,12]. In this approach, fully sampled data is subsampled and the MBDL architecture is trained to estimate ground truth data from subsampled data. Fully sampled ground truth images are impractical to acquire for many applications, especially those that require high spatial resolution dynamic acquisitions. This includes free breathing, 3D, non-Cartesian imaging as k-space sampling is limited by respiratory motion. For instance, even for longer (5-10 minute) scans, end inspiratory frames from respiratory gated 3D pulmonary UTE acquisitions contain very few projections (e.g. 2-7k projections). The sampling of these respiratory states is often not known a priori, and the acquisition time required to collect close to fully sampled frames (>100k projections) across all respiratory states would potentially take several hours.

Self-supervised MBDL[15] is a promising method that allows training without ground truth data. In this approach, k-space is partitioned into two subsets. One k-space subset is used as input to the MBDL architecture, and the second k-space subset is used only in the self-supervised loss term during training. Self-supervised MBDL is then trained to start with data from one subset of k-space data and solve for the other[16]. A recent extension of this approach called multi-mask self-supervised MBDL improves upon the original method **(citation)** by enforcing self-supervised loss simultaneously across multiple pairs of disjoint k-space subsets allowing for more efficient use of limited data with the caveat of increased training time. These methods have the potential to address the difficulty obtaining training data for 3D non-Cartesian acquisitions; however, self-supervised MBDL and its multi-mask extension leverage only spatial correlations in data (spatial self-supervised MBDL). Reconstruction methods that leverage spatial correlations alone are typically unable to recover images comparable to state-of-the-art iterative methods for highly undersampled 3D non-Cartesian datasets. Further, for multi-channel, 3D non-Cartesian data, the multi-mask extension results in unrealistic training times on the order of weeks as 3D nonuniform fast fourier transforms must be applied many times across multiple frames.

Methods used to reconstruct 3D non-Cartesian data often use the fact that acquisitions are acquired dynamically and take advantage of shared information across frames. These methods can reconstruct more highly under-sampled data than methods that rely on spatial regularization alone. Examples include temporal difference compressed sensing across respiratory states (XD-GRASP[10]), kt-PCA[17], multi-tasking [18], and low rank reconstructions[10,19]. Work incorporating non-rigid motion field estimation into these reconstructions has demonstrated even higher quality results as aligning data spatially across frames reduces temporal variations[20,21] e.g. iMoCo.

There are several recent deep learning techniques that explore incorporating spatiotemporal correlations across frames and motion correction into deep learning frameworks to improve reconstruction quality**(citation).** These works, however, use Cartesian data that can be divided into smaller subsets for efficient training and thus do not have to manage the memory or training time constraints seen when reconstructing 3D non-Cartesian data. Further, these works focus on either MBDL-based supervised training approaches **(citation)** or purely image space noise2noise approaches **(citation)** as opposed to self-supervised training. To our knowledge, the only self-supervised learning work using non-Cartesian data is limited to 2D single frame acquisitions **(citation)**.

In this work, we investigate the combination of a self-supervised learning based eXtra Dimension MBDL (XD-MBDL) architecture that takes advantage of shared information across frames with motion compensation to reconstruct a single image from free breathing 3D non-Cartesian acquisitions data-consistent with a user selected respiratory phase. We apply this technique to reconstruct the end-inspiratory phase from high resolution (1.25 mm isotropic) respiratory binned, 3D pulmonary UTE acquisitions, investigate the impact of motion correction on reconstruction quality, and compare performance to spatial self-supervised MBDL, XD-GRASP and iMoCo reconstructions. Our primary contribution is to demonstrate that the challenge of reconstructing highly undersampled frames using the original spatial self-supervised MBDL can be overcome by allowing the model to incorporate spatiotemporal correlations from other acquired frames *without altering either downstream data-consistency terms or the self-supervised loss* seen in the original spatial self-supervised model. This allows the efficient training times seen in spatial-only self-supervised approach to be preserved while significantly improved reconstruction quality for highly accelerated acquisitions.

## 2. Theory
## 2.2. Self-supervised MBDL

Consider the problem of reconstructing an image from undersampled data $y$. For highly accelerated acquisitions, this problem is ill-posed and is typically solved by minimizing a regularized least squares objective function:

$$\|Ex - d\|_2^2 + \lambda R(x) \quad [1]$$

where $x$ is the image to be reconstructed, $E$ is the non-uniform Fourier transform (NUFFT) operator including sensitivity maps, $y$ is k-space data, and $R$ is the regularizer with weight $\lambda$. The L2 loss term ensures that solutions remain consistent with the acquired data. The regularizer term constrains $x$ to satisfy certain properties resulting in removal of undersampling artifacts. Classic regularizers include sparsity in a given transform domain and nuclear norm minimization to enforce low rankness. Recent work replaces these classic handed-crafted regularizes with convolution neural networks (CNNs) that learn to remove undersampling artifact during training in a framework known as model based deep learning (MBDL).

MBDL unrolls a fixed number of iterations alternating between data consistency steps and CNN regularization steps. To accelerate convergence, MBDL architectures often use conjugate gradient iterations for data consistency[11]. The majority of work using MBDL has focused on 2D Cartesian reconstructions[11,12,23] trained using fully sampled ground truth data minimizing:

$$\min_{\theta} \frac{1}{N} \sum_{i=1}^{N} \mathcal{L}_I(x_{ref}^i, f(y_\Omega^i, E_\Omega^i; \theta)) \quad [2]$$

Where $\mathcal{L}_I()$ is the supervised loss in image space (often an L2 norm), $x_{ref}^i$ is an example ground truth image from the training set, $y_\Omega^i$ is retrospectively undersampled k-space data, $f(y_\Omega^i, E_\Omega^i; \theta)$ is the MBDL image-space output architecture with $E_\Omega^i$ representing the combination of the Fourier transform operator and coils sensitivity maps, and $\theta$ is the set of learnable network weights.

Noise2Noise (N2N)[16] methods offer an alternative approach to supervised methods when ground truth data is unavailable. In place of supervised training that uses pairs of corrupted and ground truth images to learn to remove various artifacts, Lehtinen Et al.[16] show that simply by training on pairs of differentially corrupted images, neural networks can learn the average of the distribution of these corrupted images i.e., the clean image. In the original N2N paper[16], this method was applied to MRI reconstruction as proof of concept. Fully sampled Cartesian k-space brain data was subsampled to generate undersampled image pairs from the same volume. A purely data driven network architecture without model-based data-consistency steps was trained with 5,000 image pairs by enforcing the L2 norm between the Fourier transformed neural network output and the k-space data unseen by the network. On retrospectively undersampled test data (up to 10X acceleration), N2N reconstructions had comparable PSNR and visual quality improvements to images reconstructed by networks trained using supervision. Recent work by Yaman Et al.[15] explored the performance of N2N approaches trained by subsampling accelerated k-space data and integrating the N2N framework into MBDL. Their work[15] demonstrated that this approach could still reconstruct high quality images comparable to supervised methods when trained on undersampled Cartesian data.

In self-supervised MBDL, undersampled k-space Ω is divided into disjoint subsets Θ and Λ such that Ω = Θ ∪ Λ . The supervised loss in Eq. 1 is replaced with:

$$\min_\theta \frac{1}{N} \sum_{i=1}^{N} \mathcal{L}_k(y_\Lambda^i, E_\Lambda^i f(y_\Theta^i, E_\Theta^i; \theta)) \qquad [3]$$

Where $\mathcal{L}_k$ is a self-supervised loss enforced in k-space, $y_\Lambda^i$ represents k-space entries associated with k-space subset Λ, $y_\Theta^i$ represents k-space entries associated with k-space subset $\theta$. The loss is enforced in k-space between the image space output of MBDL transformed back to k-space: $E_\Lambda^i f(y_\Theta^i, E_\Theta^i; \theta)$ and $y_\Lambda^i$. A limitation of this architecture is that it relies on spatial CNNs alone and does not harness dynamic information.

## 2.2 XD-MBDL Architecture

To extend spatial self-supervised learning to applications with multiple frames, we propose XD-MBDL, a self-supervised architecture that leverages shared information across frames to boost image quality. To incorporate the ability to leverage shared information across frames into MBDL, we propose a small modification to the original self-supervised MBDL model[15]. For the first unroll, the spatial residual network is replaced with an encoder-like residual network that takes in N data-frames along the channel dimension as input and outputs a single frame ($out_1$). This output is then passed to conjugate gradient data-consistency steps and spatial residual networks in downstream unrolls. All data-consistency steps and the self-supervised loss are enforced only on a single target frame **(figure 1a).** Incorporating this encoder-like architecture (**figure 1b**) into the MBDL architecture has several benefits. Passing from N frames to one frame acts as a bottleneck that forces information sharing along the temporal dimension. This bottleneck is critical not only for removal of undersampling artifact, but also for training and memory efficiency as downstream data-consistency layers and the self-supervised loss need only be enforced on one frame. Such memory savings are particularly relevant for 3D non-Cartesian imaging where memory demands are already at the limits of current hardware.

**Fig. 1 appears here**

XD-MBDL is trained using a self-supervised approach. T frames of k-space data vectors are organized into an array $Y^i$. Under-sampled k-space $\Omega_1$ corresponding to frame one is partitioned into disjoint subsets $\Theta_1$ and $\Lambda_1$ such that $\Omega_1 = \Theta_1 \cup \Lambda_1$. Frame one in $Y^i$ is the target frame which the self-supervised loss is enforced on during training. The undersampled k-space associated with all other frames is not partitioned. It follows then that $Y^i = [y^i_{\Theta_1}, y^i_{\Omega_2}, \ldots, y^i_{\Omega_T}]$. The k-space vectors in $Y^i$ are transformed into images using an adjoint NUFFT and used as input to the XD-MBDL architecture. The self-supervised loss for the XD-MBDL architecture then is only slightly modified from **eqn. 3** to allow for multi-frame inputs:

$$\min_{\theta} \frac{1}{N} \sum_{i=1}^{N} \mathcal{L}_k(y^i_{\Lambda_1}, E^i_{\Lambda_1} f(Y^i, [E^i_{\Theta_1}, E^i_{\Omega_2}, \ldots, E^i_{\Omega_T}]; \theta) \quad [4]$$

Where $\mathcal{L}_k$ is the self-supervised loss in k-space (L2-norm), $y^i_{\Lambda_1}$ is the vector of k-space data associated with $\Lambda_1$ from frame 1, and $Y^i$ is as defined above. The loss is enforced between the image space neural network output transformed back to k-space $E^i_{\Lambda_1} f(Y^i, [E^i_{\Theta_1}, E^i_{\Omega_2}, \ldots, E^i_{\Omega_T}]; \theta)$ and $y^i_{\Lambda_1}$ both from frame 1.

## 3 Methods
### 3.1 Overview
We applied XD-MBDL to reconstruction of the end-inspiratory phase of free breathing, retrospectively gated 3D pulmonary UTE acquisitions. End-inspiratory frame images are commonly used in CT for pulmonary nodule detection[24] and are also required for MRI ventilation mapping[25] hence our focus on this frame. Further, the end-inspiratory frame is often the most undersampled respiratory phase, and is thus challenging to reconstruct. The overall workflow can be seen in **figure 2**. Motion fields were estimated using GPU based image registration from an initial low resolution (3mm isotropic) XD-GRASP reconstruction. We rely on an initial iterative reconstruction as it eliminates concerns over accurately reconstructing datasets with complex respiratory motions that neural network approaches trained on volunteer data may not appropriately capture. Motion fields from this image registration were then applied to motion correct both the training and test data to train a XD-MBDL architecture used to reconstruct end-inspiratory phase images. In this work, k-space space data is binned into six respiratory phases.

**Fig. 2 appears here**

### 3.2 XD-MBDL Implementation

The XD-MBDL architecture was unrolled for five iterations alternating between residual CNNs (32 channels/conv, 3D conv with 3 x 3 x 3 kernels, no bias, eight convolutional layers per CNN) with ReLU activations and conjugate gradient data-consistency steps with five inner iterations. Conjugate gradient data-consistency was applied similar to that found in Aggarwal Et.al[11] with learnable parameter $\alpha$. The encoder-like network discussed in **section 2.2** was used in the first unroll while spatial residual networks were used in remaining unrolls. For the encoder-like network, N complex valued respiratory phase images were converted to 2N channel data as input. The output from the data-sharing network was a 2-channel image that was then passed to subsequent data-consistency and spatial residual networks. The complex valued volume output from each data-consistency step was converted to 2-channel data as input for each spatial residual network. The self-supervised loss was implemented as an L2 norm in k-space summed over channels.

Block-wise learning with gradient checkpointing was used for memory efficient XD-MBDL training[13]. Without this method, reconstruction of these high resolution, volumetric datasets would be difficult even on GPU clusters. For a single unroll, this technique decomposes input volume(s) into patches, checkpoints each patch, iteratively passes these patches through the network, and then recombines the output patches into the full volume for data-consistency. Each input volume was decomposed into eight patches. Gradient checkpointing was also applied to the multi-channel data-consistency step to reduce memory use.

### 3.3 Motion Compensation Workflow
**Non-Cartesian Data Acquisition and Retrospective Respiratory Binning**

The University of Wisconsin IRB approved all study procedures and protocols for volunteer and clinical studies following the policies and guidance established by the campus Human Research Protection Program. All study procedures were performed according to the Declaration of Helsinki, including obtaining written informed consent from each participant. Post-Ferumoxytol (4mg/kg) Free Breathing UTE lung acquisitions and Free Breathing UTE lung acquisitions without contrast in healthy volunteers were used for training and testing. 16 cases were included with 7 cases used for training, 1 case for validation, and 8 cases for testing. Datasets were acquired with a 3T GE Scanner with a 32-channel coil. Scan parameters for the Post-Ferumoxytol data included scan time of 5:45minutes, TE=0.25ms, TR=3.6ms, and 1.25mm isotropic resolution. A total of 94,957 projections were acquired using 3D pseudorandom bit-reversed view ordering with readout length of 636 points[1] per acquisition. Similar scan parameters were used for the non-contrast data, however, the TE=80$\mu$s was shorter. Respiratory positions were recorded with a respiratory belt. Data was coil compressed to 20-channels using PCA coil compression[27]. The acquisition provided whole chest coverage with matrix sizes varying between 300-450 x 200-300 x 300-450 based on automatic field of view determination at full resolution. Density compensation was normalized using the max eigenvalue of the NUFFT operator and k-space was then rescaled based on this value [6]. Respiratory belt signal was used to divide data into six respiratory phases from end-inspiration to end-expiration. Bins were set by the respiratory signal such that each bin was allowed to have an unequal number of projections.

**Motion Field Estimation**

XD-GRASP was used to reconstruct low resolution (3mm isotropic) motion resolved images that were used to estimate motion fields. GPU-based image registration was performed by modifying the algorithm proposed in recent work[22]. In the original registration method, motion fields through time are represented as multi-scale low rank left spatial and right temporal bases estimated by warping a template frame with loss enforced in k-space. Here we still estimate motion fields as multi-scale low rank components but replace template frame warping and k-space loss with more traditional warping of respiratory phase images to a reference state and loss enforced in image-space. See **supplement 7.2** for more details.

For implementation, end-inspiration was chosen as the reference phase for motion correction as this eliminated the need to estimate a second set of motion fields that warp registered data back to the motion state for data-consistency. L2 norm was used for the similarity metric with no explicit regularization applied. 30 epochs of the registration algorithm at lower resolution were run with each epoch consisting of stochastic updates over all frames. Once initial lower resolution fields were estimated, the motion fields were interpolated to the desired resolution. This method was implemented using auto-differentiation in Pytorch with an Adam optimizer with learning rate of .01. Motion fields were estimated for both training and test data.

**Motion Compensated XD-MBDL**

   **Training:** XD-MBDL was trained on registered data with end-inspiration as the target phase. Training input was generated as follows:

1. Prior to training, gridded respiratory phase images were generated from all respiratory phases except end-inspiration (the target frame) without partitioning k-space. This data did not change over iterations, so it only needed to be computed once.

For each training iteration:

2. End-inspiratory phase k-space data was randomly partitioned along the radial dimension into two disjoint subsets such that $\Omega_{end-insp} = \Theta_{0.6} \cup \Lambda_{0.4}$ where $\Theta_{0.6}$ represents 60% of the radial spokes, $\Lambda_{0.4}$ represents 40% of the radial spokes. This partition was chosen based on the results from Cartesian data in Yaman Et. al [15].
3. Gridded end-inspiratory images were generated from the k-space data corresponding to $\Theta_{0.6}$.
4. Images created in step 1 were stacked with the gridded end-inspiratory image into an array, registered and then used as input into MBDL.
5. $\Lambda_{0.4}$ was used only in the self-supervised loss

This architecture was implemented in Pytorch[28] for 2000 iterations using an Adam optimizer with learning rate of 1e-3 and NUFFTs from SigPy[29] on Intel Xeon workstations using one 40 GB A100 GPU.

**3.4 Evaluation**

   Image quality was assessed similar to the approach used by Zhu Et. al.[20]. Apparent signal to noise (aSNR), defined as the signal in a region of interest divided by the standard deviation of signal outside the body, was measured in the aortic arch, parenchyma, and airway. Contrast to noise ratio (CNR), defined as the contrast difference between a selected region of interest versus the standard deviation of signal outside of the body, was measured between the aorta and airway and parenchyma and airway. Quantitative metrics across various reconstructions were compared using paired t-testing. Differences between reconstructions were considered significant if $P < .05$.

   The impact of motion correction on reconstructed image quality was investigated by training separate XD-MBDL architectures on unregistered and registered data. Both architectures were then subsequently tested on unregistered and registered data. This analysis was limited to Ferumoxytol data.

   The ability of XD-MBDL to generalize to different contrasts was evaluated by training three separate architectures on Ferumoxytol data only, non-Contrast data only, and both contrasts respectively.

Reconstruction quality on both Ferumoxytol and non-Contrast data using each architecture was then assessed.

Finally, XD-MBDL end-inspiratory image quality and run-time were compared to spatial self-supervised MBDL, XD-GRASP, and iterative motion compensated reconstructions (iMoCo) for both Ferumoxytol and non-Contrast data. The XD-MBDL architecture trained on data that generalized best based on both visual quality and quantitative metrics to all data-sets was used for this comparison .CG-SENSE was used as baseline. Spatial self-supervised MBDL was trained on end-inspiratory data using the same training data and k-space partitioning used for XD-MBDL. The spatial self-supervised MBDL architecture, however, used only spatial residual CNNs for all unrolls. XD-GRASP reconstructions were implemented by modifying the code available at https://github.com/mikgroup/extreme_mri/ to accept previously binned data based on respiratory belt signal as input. Reconstructions were run for 200 iterations with a temporal difference regularization weight of 1e-6. iMoCo reconstructions were performed using modified code from https://github.com/PulmonaryMRI/imoco_recon to allow bins with varying number of spokes. The ANTS CPU based image registration used in this code was replaced with the GPU-based registration used in this work to reduce computation time. As iMoCo reconstructs a template image data-consistent with all respiratory phases, we aligned this template image with the end-inspiratory phase for comparison to XD-MBDL. Eighteen iterations of iMoCo were run. CG-SENSE was run for 20 iterations

# 4 Results
## 4.1 Impact of Motion Correction during Training and Inference on XD-MBDL Image quality

**Fig. 3 Appears Here**

**Figure 3** shows sagittal end-inspiratory slices for combinations of XD-MBDL architectures with training/inference on registered/unregistered Ferumoxytol data. XD-MBDL trained on unregistered data resulted in improved image quality relative to CG-SENSE, and this improvement was observed independent of whether the data was registered during inference. Reconstruction quality was further improved when input data was registered during both training and inference (Train:reg/Inf:Reg is motion compensated XD-MBDL). Motion compensated XD-MBDL reconstruction had minimal streaking artifact and resolved even small vascular features nicely (orange arrow). Training XD-MBDL on registered data and then performing inference on unregistered data (Train: Reg/Inf: Unreg), however, led to significant feature distortion suggesting this architecture did not learn to preserve motion state. In fact, it appears that the liver edge in this reconstruction (**figure 3, column 4**) was biased towards end-expiration. Notably, this effect was not present in the architecture with training and inference on unregistered data.

**\*\*Fig. 4 appears here\*\***

**Figure 4** shows aSNR and CNR values across these same reconstructions over all Ferumoxytol cases in the test dataset. Motion Corrected XD-MBDL had significantly higher aortic arch, parenchyma, and airway aSNR values than CG-SENSE and XD-MBDL models trained with unregistered data (**Figure 4a**, aorta/parenchyma: *P<.001* , airway: *P<.05*). Although airway aSNR should be close to zero, the CNR (**Figure 4b**, *P<1e-3*) for the parenchyma and aortic arch (relative to the airway) remained higher for motion-compensated XD-MBDL than these other reconstructions suggesting that these structures were better distinguished using motion compensated MBDL.

Although XD-MBDL trained on registered data with inference on unregistered data had similar aSNR (aorta arch: *P<.312*, parenchyma: *P<.332*, airway: *P<.074*) and CNR (aortic arch: *P< .2788*, parenchyma: *P<.262*) compared to motion compensated XD-MBDL, substantial reconstruction error was present (**figure 3**).

### 4.2 Ability of XD-MBDL to generalize to Different Contrasts

**\*\*Fig. 5 appears here\*\***

**Figure 5** shows example non-contrast (top row) and Ferumoxytol (bottom row) reconstructions for XD-MBDL architectures trained on mixed contrast data, non-contrast data only and Ferumoxytol data only respectively. Visually, reconstruction quality for both example cases appears to be nearly independent of the data the architecture was trained on as all reconstructions are significantly improved over the CG-SENSE baseline. No statistically significant differences were seen for either aSNR (aorta, airway, and parenchyma: *P>.05* ) or CNR (aorta and parenchyma: *P>.05*) across training approaches.

**XD-MBDL Image quality Comparison**

**Figure 7**6show coronal slices for motion compensated XD-MBDL, iMoCo, XD-GRASP, Spatial Self-supervised MBDL, and CG-SENSE reconstructions for Ferumoxytol test data. From **figure 6,** motion compensated XD-MBDL has improved image quality over XD-GRASP, spatial self-supervised MBDL, and CG-SENSE with improved ability to resolve small vascular features. Relative to iMoCo, motion compensated XD-MBDL does have sharper features (orange arrow).

**Fig. 6 appears here**

**Figure 7** similarly shows coronal slices for motion-compensated XD-MBDL, iMoCo, XD-GRASP, Spatial Self-supervised MBDL, and CG-SENSE reconstructions for non-contrast test data. Motion compensated XD-MBDL again demonstrates improved image quality over XD-GRASP, spatial self-supervised MBDL and CG-SENSE. In this example case, motion compensated XD-MBDL clearly resolves smaller vascular features better than iMoCo.

**Fig. 7 appears here**

For the Ferumoxytol training data, motion compensated XD-MBDL had significantly higher aorta and parenchyma aSNR ($P<.001$) than all other reconstructions. Motion compensated XD-MBDL airway aSNR was significantly higher than CG-SENSE ($P<.05$); however, it did not differ significantly from spatial self-supervised learning ($P<.26$), iMoCo ($P<.07$) or XD-GRASP ($P<.434$) **(Figure 8a).** Both aorta ($P<.01$) and parenchymal CNR ($P<.01$) were significantly higher for motion compensated XD-MBDL than all other reconstructions **(Figure 8b).**

**Fig. 8 appears here**

For the non-contrast training data, motion compensated XD-MBDL has significantly higher aorta and parenchyma aSNR ($P<.05$) than all other reconstructions except for iMoCo. No significant differences in any aSNR values were seen between motion compensated XD-MBDL and iMoCo **(Figure 9a and Figure 9b)**

**Fig. 9 appears here**

Detailed run times for implementations of the various reconstruction methods can be found in **supplemental section 7.1** The entire motion compensated workflow during inference required ~15 minutes of runtime during inference. The motion resolved reconstruction and image registration took up much of the reconstruction time (12 min). The overall XD-MBDL workflow was substantially shorter (~13 minutes) than the reconstructions for XD-GRASP (~32 minutes) and iMoCo (~42 minutes), but longer than spatial self-supervised MBDL (~1 minute). Notably, spatial self-supervised MBDL and XD-MBDL had nearly identical per iteration training times.

## 5 Discussion

This work investigated the combination of a new self-supervised MBDL architecture called XD-MBDL and a GPU based image registration technique to develop a motion compensated DL framework. The goal of this work was to show that the self-supervised model proposed by Yaman Et. Al. could be adapted to the highly undersampled, high dimensional setting simply by incorporating spatiotemporally correlated frames while still only enforcing data-consistency and the self-supervised loss on one frame of interest allowing for memory and time efficient training.

This method was used to reconstruct highly undersampled, end-inspiratory images from respiratory binned, free breathing, 3D Pulmonary UTE acquisitions. This technique (during inference) consisted of a respiratory binned acquisition, motion field estimation from a low resolution XD-GRASP motion resolved reconstruction, and a final motion compensated XD-MBDL step. We demonstrated that incorporating motion compensation into XD-MBDL improved the quality of these deep learning reconstructions (**figure 3**). We then showed that XD-MBDL was able to generalize well to test data with very different contrast characteristics from the data the architecture was trained on **(figure 5)**. Finally, we demonstrated that our motion compensated method resulted in higher quality images than spatial self-supervised deep learning or XD-GRASP (**figure 6 and** f**igure 7).** For Ferumoxytol data, motion compensated XD-MBDL outperformed iMoCo both visually (**figure 6)** and based on aSNR and CNR metrics **(figure 8).** For non-contrast enhanced data, motion compensated XD-MBDL appears to resolve small features better than iMoCo, (**figure 7)** however, no difference in aSNR and CNR metrics was seen. (**figure 9).** We note that XD-MBDL was significantly faster than either XD-GRASP (32 minutes) or iMoCo.(42 minutes)

This work addresses the difficulty obtaining fully sampled 3D non-Cartesian data for supervised MBDL training by leveraging self-supervised learning. The challenge is the image quality associated with straightforward application of spatial self supervised MBDL to these highly accelerated acquisitions is not on par with state-of-the-art iterative methods like iMoCo. This can be seen in **figure 6** and **figure 7** where the spatial self-supervised reconstruction has residual undersampling artifact and loss of small vascular features while iMoCo is able to clearly resolve these vascular features. Spatial self-supervised MBDL and iMoCo use substantially different amounts of data during reconstruction which likely drives these differences in reconstruction quality. iMoCo leverages all data acquired during the scan while spatial self-supervised MBDL uses only a single, often highly undersampled, frame. Spatial self-supervised MBDL is thus data-starved relative to iMoCo.

The XD-MBDL architecture proposed here addresses the data starvation seen in spatial self-supervised MBDL by leveraging shared information across multiple respiratory frames. This architecture alone, however, was not enough to match the image quality of iMoCo primarily due to blurring of small vascular features similar to that seen in XD-GRASP (**figure 3)**. The power of sharing information across frames truly becomes apparent when combined with motion compensation. It is this combination of XD-

MBDL with motion compensation that allows our proposed method to compete with state-of-the-art iterative methods.

Unsurprisingly, motion compensated XD-MBDL shares some of the limitations seen with state-of-the-art motion corrected iterative methods (e.g. iMoCo). Both motion compensated DL and iterative approaches depend on high quality XD-GRASP motion resolved reconstructions that appropriately model respiratory dynamics and image registration approaches that accurately estimate motion fields. The need for this intermediate image reconstruction step followed by image registration does lengthen run-time. Several works **(citations)** address this issue by replacing iterative motion-resolved reconstruction and estimation with neural network steps that can rapidly reconstruct the motion resolved dataset and estimate motion.

A general limitation though of using DL based reconstructions and registration that attempt to estimate motion dynamics is the architecture's ability to appropriately model motion dynamics during inference is dependent on the similarity between motion dynamics in the test and training data. The training and test data used in this work was from healthy volunteers with periodic breathing, where the number of radial spokes binned from end-inspiration to end-expiration increased monotonically. Respiratory patterns can be highly irregular, particularly in patients with diffuse lung disease, resulting in widely varying radial spokes assigned to each respiratory phase. It is an open question how well DL methods trained on healthy volunteers with periodic respiratory patterns would generalize to such challenging datasets. Iterative approaches then although slower are more practical.

There are several limitations then to the present study that warrant further investigation. First and foremost, the lack of a ground truth made both image quality evaluation and determining whether the architecture accurately represented underlying anatomy challenging. Further, this study focused on healthy volunteers with normal respiratory dynamics potentially limiting the generalizability of our results. A primary focus of work moving forward will be on evaluating reconstructions on both greater numbers of patients and patients with a wider variety of respiratory dynamics, particularly in clinical cases where gold standard images may be available (e.g., CT) for comparison.

Another limitation of our work is the architecture was likely undertrained due to the limited number of training iterations (2000) used to keep training times reasonable. Additionally, parameters such as the number of unrolls, optimal Θ and Λ splitting ratios, and network architecture were not investigated at this stage. Finally, more work is needed to investigate the generalization of this technique to other dynamic applications, including 4D-Flow and DCE-MRI.

# 6 Conclusions

In this study, we developed a motion-compensated self-supervised MBDL reconstruction method that combines motion estimation with an MBDL architecture that leverages correlations across frames. We demonstrate on healthy volunteers that this approach allows for fast and high-quality 3D pulmonary UTE reconstructions.

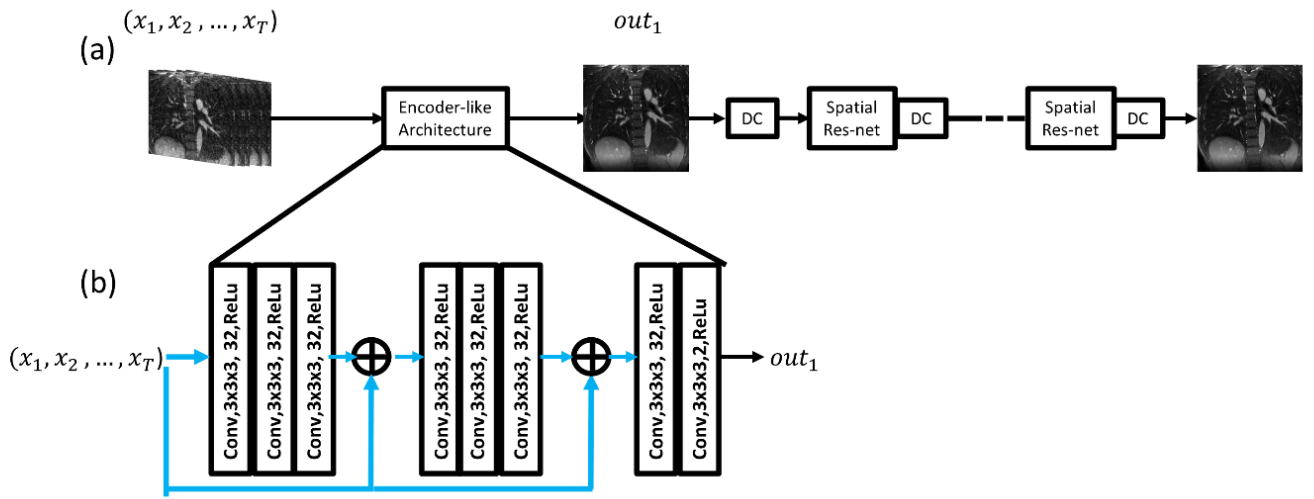

**_Figure 1_**: _XD-MBDL Architecture. An array of images_: $(x_1, x_2, ..., x_T)$ _reconstructed using a simple adjoint NUFFT is used as input to **(a)** the XD-MBDL architecture with the first frame $x_1$ chosen as the target frame for reconstruction. All data-consistency steps and the self-supervised loss are enforced on this target frame. The target frame $x_1$ is reconstructed from data corresponding to a subset $\Theta_1$ of k-space $\Omega_1$ where $\Theta_1 \cup \Lambda_1 = \Omega_1$ and $\Lambda_1$ is used solely to enforce the self-supervised loss. Gridded images $x_i$ where $i \neq 1$ are reconstructed using all available k-space data corresponding to that frame. This array is passed through **(b)** the encoder-like architecture compressing T frames to 1 frame. This output frame $out_1$ is then passed to subsequent conjugate gradient data-consistency steps and spatial residual networks (Spatial Res-net) to produce a single reconstructed image corresponding to the target image._

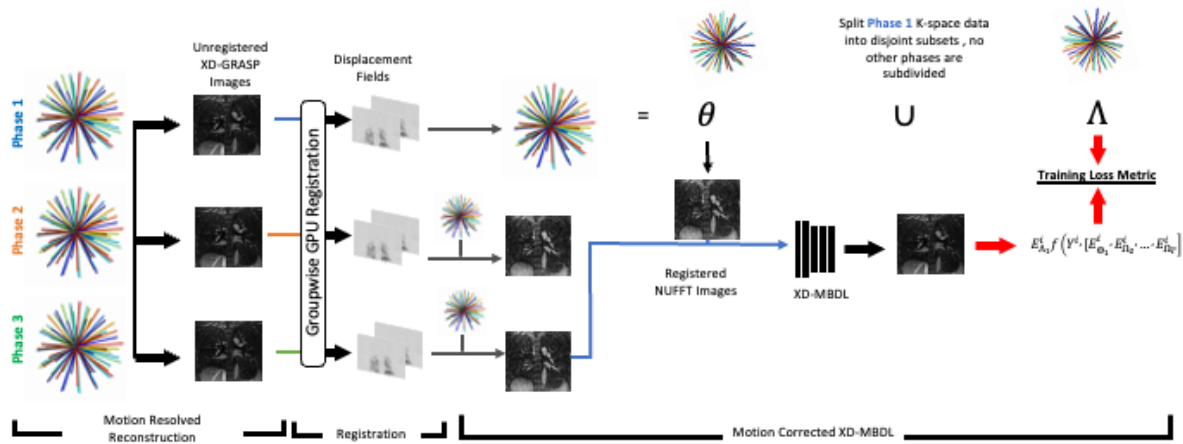

***Figure 2:*** *Motion Compensated Workflow.* ***3D Pulmonary UTE data*** *was acquired and binned based on a respiratory belt signal into N different respiratory phases. XD-GRASP was used for motion resolved reconstruction.* ***Displacement fields*** *were then estimated using GPU-based non-rigid registration with the end-inspiratory phase as reference. This registered training data was used to train a **motion compensated XD-MBDL architecture.** The architecture was trained by splitting k-space data from the end-inspiratory phase (phase 1) into two disjoint subsets θ and Λ. The gridded image from θ subset was registered and stacked with the remaining phases and then passed through the architecture. The output of XD-MBDL was then transformed back to k-space and self-supervised loss was enforced between this output in k-space and the k-space subset associated with Λ.*

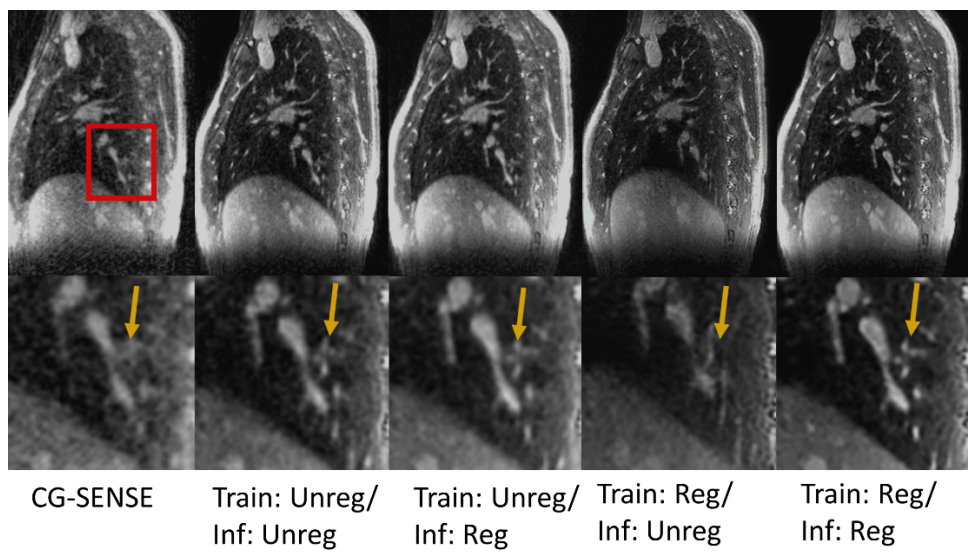

***Figure 3:*** *Impact of Registration on Reconstruction Results. Two XD-MBDL architectures were trained, one on registered data, the other on unregistered data. Reconstructions using registered and unregistered test data were then run on both architectures. Displayed here are sagittal slices from reconstructions using these different strategies. The red bounding box shows the location where the image has been zoomed in on row 2. Motion compensated XD-MBDL (Reg Train/Reg Inf.) was significantly sharper and resolved more features (orange arrow) than all other reconstructions suggesting that motion correction significantly improves reconstruction quality relative to reconstructions on unregistered data. Motion compensated XD-MBDL also remains much sharper than the architecture trained on unregistered data, but with inference on registered data (Unreg Train/Reg Inf.). Interestingly, the architecture trained on unregistered data preserved features when performing inference on unregistered data significantly better than the architecture trained on registered data with inference on unregistered data.*

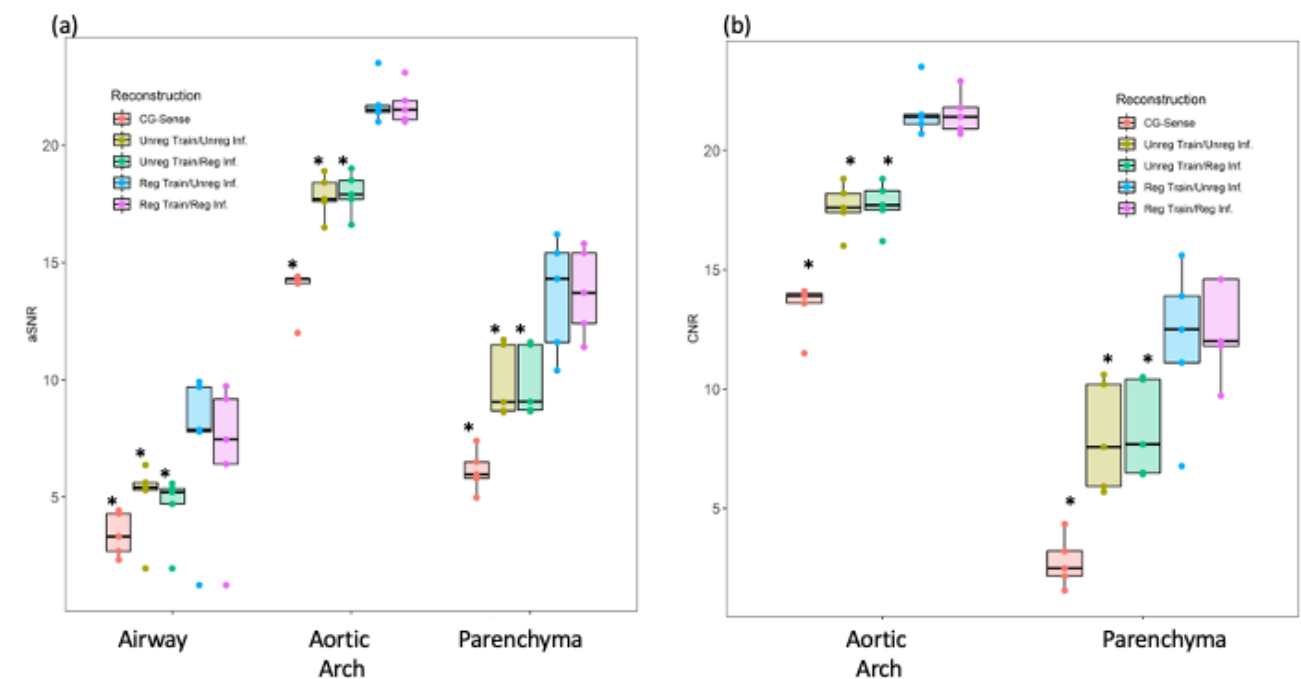

***Figure 4:*** *aSNR and CNR comparison across XD-MBDL Reconstructions. An asterisk means there is a significant difference in the metric between that architecture and motion compensated XD-MBDL (Reg Train/Reg Inf). Following Zhu Et. al: a major airway, the aortic arch, and a section of lung parenchyma were segmented in all test cases. In **(a)**, motion-compensated XD-MBDL had significantly higher aortic arch and parenchyma aSNR than all other reconstructions besides the architecture trained on registered data, but with inference on unregistered data. Note though that the image quality of this reconstruction from figure 3 was significantly worse than motion-compensated XD-MBDL. Airway aSNR should be close to 0. Here, airway aSNR was significantly higher for motion-compensated MXD-BDL than the architecture trained on unregistered data as well as CG-SENSE. However, in **(b)**, the CNR of both the aorta and parenchyma were significantly higher for motion-compensated XD-MBDL suggesting that aorta, parenchyma, and airway could be best distinguished using this reconstruction approach.*

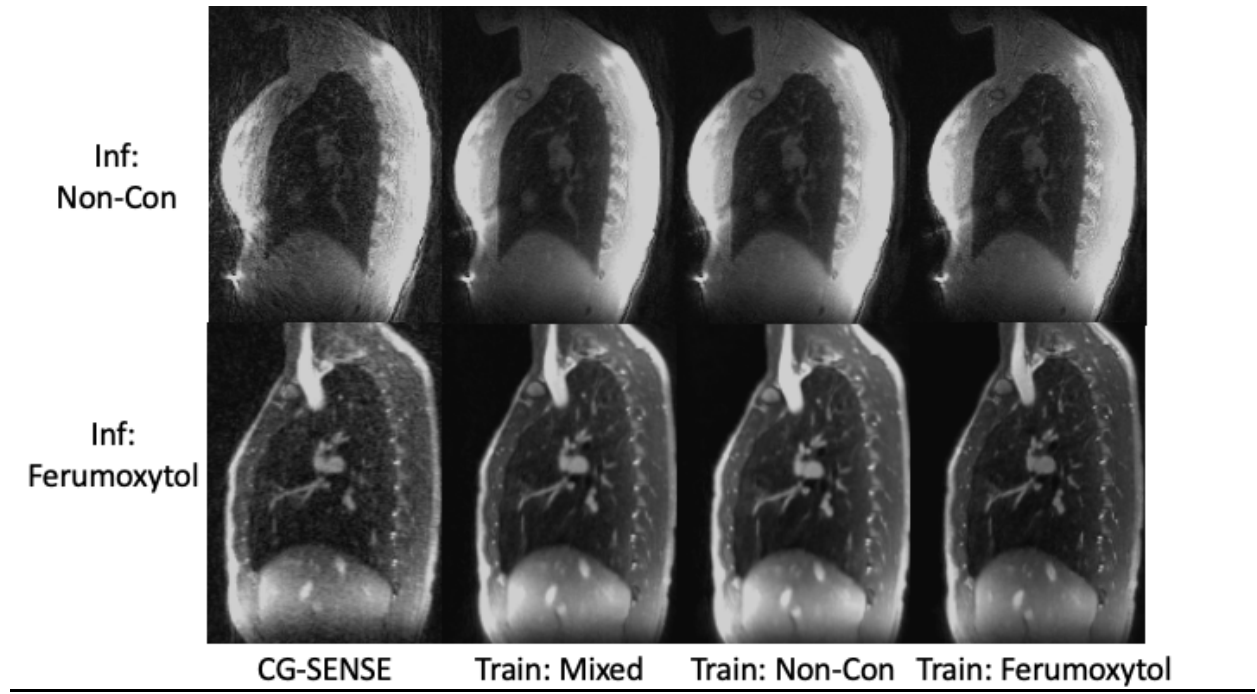

***Figure 5:**** XD-MBDL architectures trained using non-con only, Ferumoxytol only, and mixed datasets applied to non-con and Ferumoxytol data. All training approaches are visually similar in reconstruction quality except the Ferumoxytol trained, Ferumoxyotol reconstruction does appear somewhat sharper than the other reconstructions. All XD-MBDL reconstructions outperform CG-SENSE*

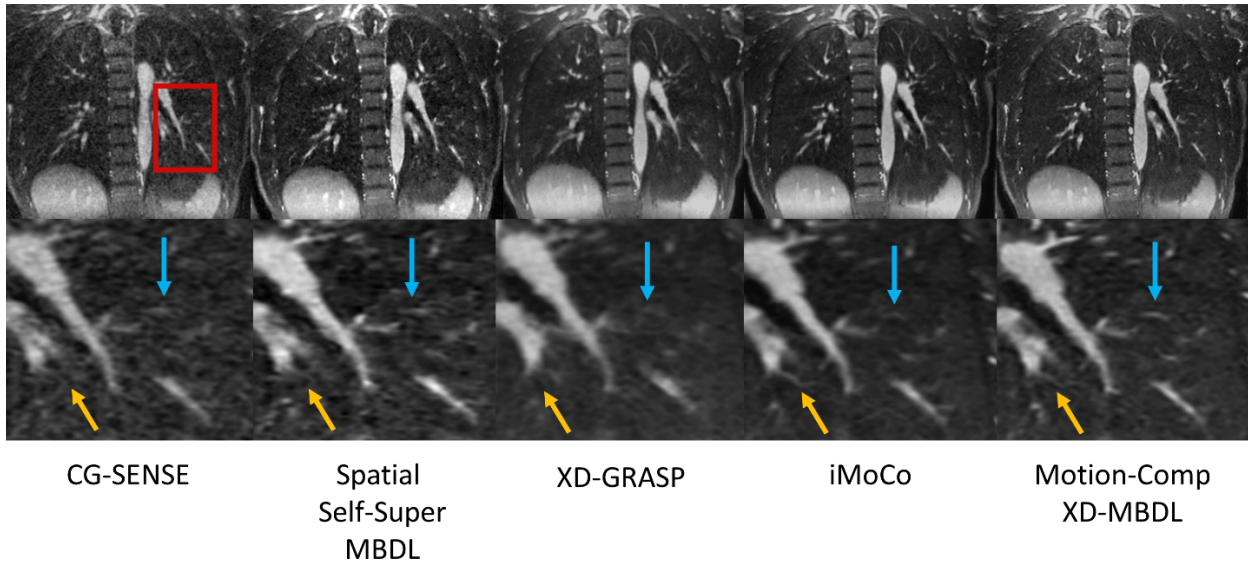

*Figure 6:* End-inspiratory Ferumoxytol reconstruction comparisons: CG SENSE vs. Spatial Self-Supervised MBDL vs XD-GRASP vs iMoCo vs. Motion compensated XD-MBDL. Displayed here are representative coronal slices from different reconstructions on the same case. The red bounding box shows the location where the image has been zoomed in on row 2. Motion compensated XD-MBDL was sharper than all other reconstructions including spatial self-supervised MBDL which had significant remaining undersampling artifact. Motion compensated XD-MBDL and iMoCo both resolved small vascular features that could not clearly be seen in the other reconstructions. iMoCo resolved some of these features (yellow arrow) more clearly than Motion compensated XD-MBDL. The reverse was also true (blue arrow).

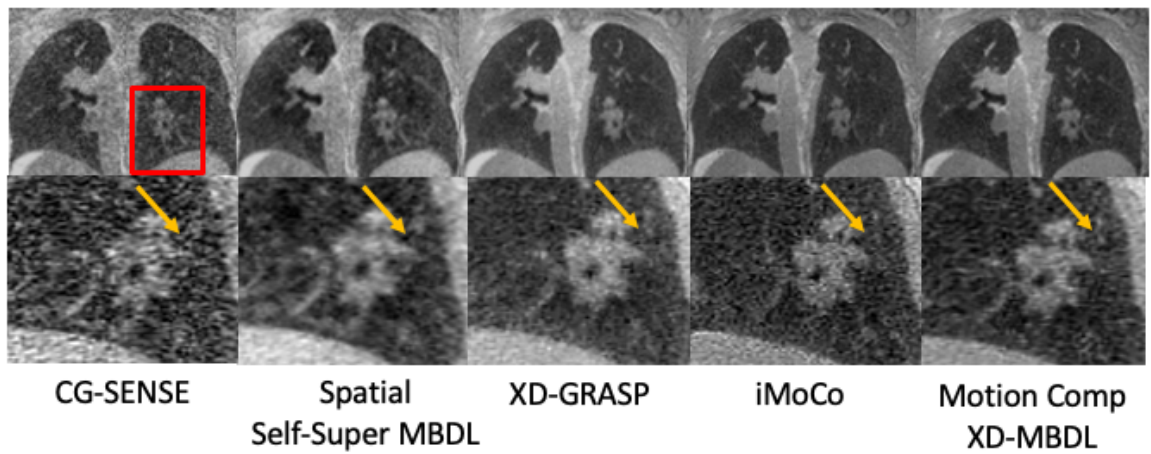

*Figure 7:* End-Inspiratory non-contrast reconstruction comparisons: CG SENSE vs. Spatial Self-Supervised MBDL vs XD-GRASP vs iMoCo vs. Motion compensated XD-MBDL. Displayed here are

*representative coronal slices from different reconstructions on the same case. The red bounding box shows the location where the image has been zoomed in on row 2. Motion compensated XD-MBDL is able to resolve small vascular features better than the remaining reconstructions (orange arrow)*

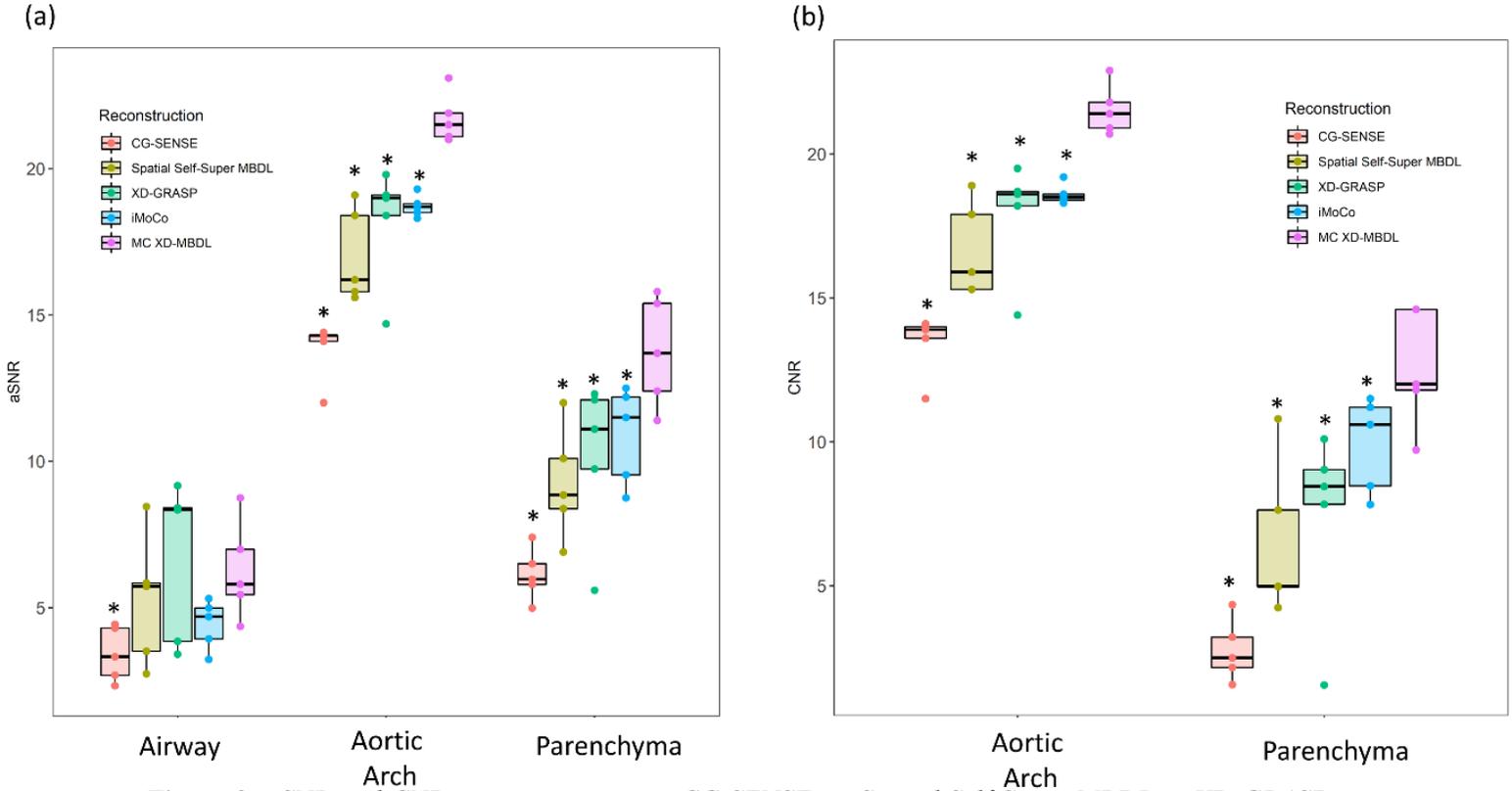

***Figure 8:*** *aSNR and CNR comparison across CG-SENSE vs. Spatial Self-Super MBDL vs XD-GRASP vs iMoCo vs. motion compensated XD-MBDL for Ferumoxytol data. An asterisk means there is a significant difference in the metric between a given reconstruction approach and motion compensated dynamic MBDL. Motion compensated XD-MBDL had significantly higher aortic arch and parenchymal aSNR as well as CNR than all other reconstructions.*

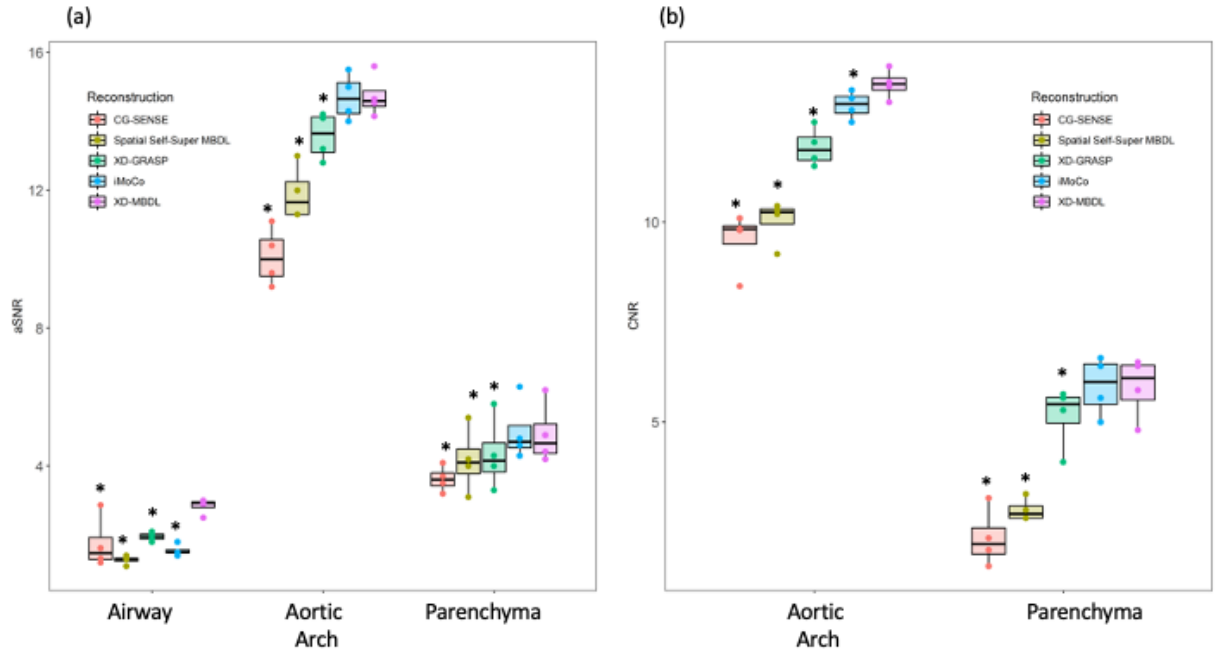

**Figure 9** *aSNR and CNR comparison for non-contrast data across CG-SENSE vs. Spatial Self-Super MBDL vs XD-GRASP vs iMoCo vs. motion compensated XD-MBDL. An asterisk means there is a significant difference in the metric between a given reconstruction approach and motion compensated XD-MBDL. No significant aSNR or CNR differences were seen between iMoCo and motion compensated XD-MBD*

**7 Supporting Information**

**7.1: Run Time Comparisons:**

**iMOCO:**

**Total Run-Time**: 42 minutes with XD-Grasp run at low resolution

**Run-Time Breakdown**

a. Motion Resolved Reconstruction low res: 6 min

b. GPU based Registration: 6 minutes

c. Final motion compensated iterative reconstruction: 30 minutes

**XD-Grasp**: 32 minutes at full resolution

**Motion Compensated Workflow**

**Total Run-time:** 13 minutes

**Run-Time Breakdown:**

a. Motion Resolved Reconstruction low res: 6 min

b. GPU based Registration: 6 minutes

c. Final motion compensated XD-MBDL: 1 minute

**Run-Time during Training/Testing:**

Forward pass: ~60 seconds

Backward pass: 205 seconds

Inference: ~60 seconds

**Spatial Self-Supervised MBDL:**

**Run-Time during Training/Testing:**

Forward pass: ~60 seconds

Backward pass: 202 seconds

Inference: ~60 seconds

### 7.3 Motion Correction Method

Incorporating motion compensation into reconstruction has been previously shown to improve image quality, and we hypothesized that the same would be true for the XD-MBDL architecture[20,21]. For motion compensation, we apply a method inspired by Ong Et al. and Huttinga Et al.[6,30], and detailed more completely in recent work (reference). In this motion correction approach, motion fields are estimated directly as multi-scale low rank (MSLR) components to regularize the registration problem along the time dimension.

Let $\phi_t \in \mathbb{R}^{3xN}$ represent a dense 3-channel deformation field of size $N$ with each voxel assigned a displacement: $\text{Id} + r(x, y, z)$ that warps a given motion state at time $t$ to a reference image. For $T$ frames, deformation fields are stacked into a spatiotemporal matrix $\Phi \in \mathbb{R}^{3xNxT}$. This matrix can be decomposed into the sum of 3-channel, rank 1, block-wise matrices across varying block scales. If $J$ is the number of block scales for the MSLR decomposition then for a given block scale $j \in J$, there are $B_j$ blocks of size $3xN_j \, x \, T$ which are then factored into a block-wise left spatial deformation field bases $\Phi_j \in \mathbb{R}^{3xN_jx1}$ and right temporal bases: $\Psi_j \in \mathbb{R}^{3xTx1}$. The sum of this decomposition across block-sizes for $\phi_t$ is:

$$\Phi_t = \sum_{j=1}^{J} M_j(\Phi_j \Psi_{j,t}^H) \quad [1]$$

Where $M_j$ is the block to 3 channel deformation field operator. This representation can then be incorporated into the classic registration problem:

$$\min_{\substack{\Phi_j, \Psi_j \\ \forall j \in J}} \mathcal{L}\left(I_{ref}, I_t(\phi_t)\right) \quad [2]$$

Where $\mathcal{L}()$ is restricted to be a pair-wise loss in this work, $I_{ref}$ is the selected reference image, $I_t$ is the motion state to be warped, and $\Phi_t$ is the motion field such that $\Phi_t = \sum_{j=1}^{J} M_j(\Phi_j \Psi_{j,t}^H)$.